\documentclass[prl,twocolumn,amsmath,amssymb,groupedaddress]{revtex4}

\newcommand{\s}{{\sigma}}

\newcommand{\xv}{{\bf x}}
\newcommand{\kv}{{\bf k}}
\newcommand{\eps}{{\varepsilon}}

\usepackage[dvips]{graphicx}

\begin{document}
\title[]{Superfluid transitions in bosonic atom-molecule mixtures near
Feshbach resonance}
\author{Leo Radzihovsky}
\author{Jae Park}
\affiliation{Department of Physics, University of Colorado,
Boulder, CO 80309}
\author{Peter B. Weichman}
\affiliation{ALPHATECH, Inc., 6 New England Executive Place,
Burlington, MA 01803}


\date{\today}

\begin{abstract}
We study bosonic atoms near a Feshbach resonance, and predict that
in addition to a standard normal and atomic superfluid phases,
this system generically exhibits a distinct phase of matter: a
molecular superfluid, where molecules are superfluid while atoms
are not. We explore zero- and finite-temperature properties of the
molecular superfluid (a bosonic, strong-coupling analog of a BCS
superconductor), and study quantum and classical phase transitions
between the normal, molecular superfluid and atomic superfluid
states.
\end{abstract}

\maketitle

Experimental realizations and coherent manipulation of
trapped degenerate gases \cite{JILAbec,MITbec} is leading to
exciting possibilities for studies of quantum liquids in
previously unexplored (e.g., extremely coherent and
nonequilibrium) regimes.  Magnetic field-induced Feshbach
resonance (FBR) in ultracold atom collisions allows fine
tuning of interactions in these quantum fluids, and was recently
used to create a degenerate mixture of coherently-coupled alkali
atoms and their diatomic molecules \cite{donley}.

\begin{figure}[bth]
\centering
\setlength{\unitlength}{1mm}
\begin{picture}(70,65)(0,0)
\put(-30,-62){\begin{picture}(60,60)(0,0)
\includegraphics{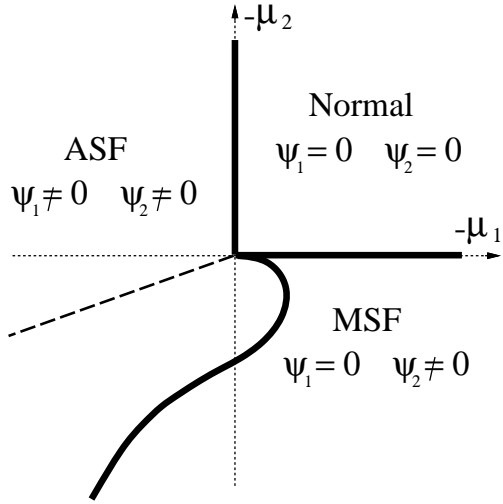}
\end{picture}}
\end{picture}
\caption{Mean-field phase diagram for a bosonic atom-molecule mixture,
showing molecular (MSF) and atomic (ASF) superfluid phases.}
\label{phasediagramA}
\end{figure}

\begin{figure}[tbh]
\centering
\setlength{\unitlength}{1mm}
\begin{picture}(70,50)(0,0)
\put(-22,-66){\begin{picture}(50,50)(0,0)
\includegraphics{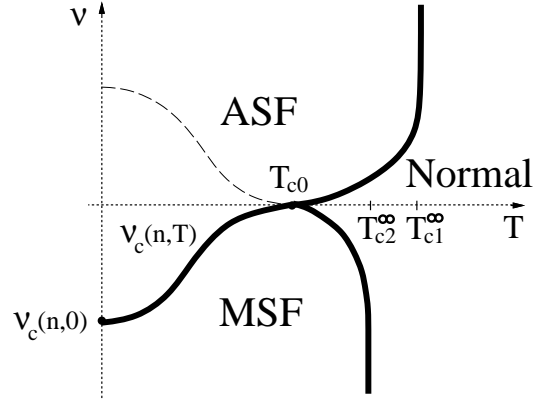}
\end{picture}}
\end{picture}
\caption{Phase diagram for a bosonic atom-molecule mixture in
$d=3$, expressed in terms of 
detuning $\nu$ and temperature $T$. It illustrates a finite-T
tricritical point at $T_{c0}$ and  a quantum critical point at
$\nu_c(n,0)$. In the weakly-interacting limit appropriate to
experiments the ratio $T_{c1}^\infty/T_{c2}^\infty = 2^{5/3}$.}
\label{phasediagramB}
\end{figure}

In this Letter we study phases and phase transitions that take
place in bosonic atom-molecule mixtures. Our main contribution is the
prediction of a thermodynamically distinct ``molecular
superfluid'' (MSF) phase, that, as illustrated in 
Figs.\ \ref{phasediagramA}, \ref{phasediagramB} ubiquitously
intervenes between the ``normal'' (N) and ``atomic superfluid''
(ASF) phases. Molecular superfluidity [and accompanying
off-diagonal long-range molecular order (ODLRO)] distinguishes MSF
from the normal state, and the {\em absence} of atomic
superfluidity from the ASF, in which both bosonic atoms and
molecules display ODLRO.  If atomic and molecular components can
be imaged independently \cite{jinBCS}, in a harmonic trap MSF
should be easily identifiable by a sharp Bose-Einstein
condensation (BEC) peak in the molecular density profile and a
broad, seemingly normal, thermal atomic cloud.

As a conventional superfluid, MSF is characterized by a
(molecular) acoustic second-sound mode. However, MSF also exhibits
a {\em gapped}, Bogoliubov-like mode, derived from unpaired atom
excitations.  MSF ground state (bosonic analog of the BCS state)
exhibits strong (atom and molecule) pairing correlations that in a
trap should be observable in the atomic density-density
correlation function. Experimentally, MSF should be accessible by
tuning temperature, atomic density (or number), and detuning
$\nu$. The MSF-ASF transition is in the $(d+1)$- and
$d$-dimensional Ising universality classes for $T=0$ and finite
$T$, respectively, and is {\em reentrant} as a function of
detuning $\nu$ and density $n$. The tricritical point, where N,
MSF and ASF meet, exhibits nontrivial and, to our knowledge,
unexplored quantum critical behavior for $d < 4$. We now sketch
derivation of these results.

Near a FBR a bosonic atom-molecule system is characterized by the
grand-canonical Hamiltonian $\hat{H}_\mu = \hat{H} - \mu \hat{N}$
\cite{foot1}
\begin{eqnarray}
\hat{H}_\mu &=& \int d^d{\bf x} \Bigg[\sum_{\s=1}^2
\left(\hat{\psi}^\dagger_\s \hat{h}_\s \hat{\psi}_\s
+ \frac{g_\s}{2} \hat{\psi}^\dagger_\s
\hat{\psi}^\dagger_\s \hat{\psi}_\s \hat{\psi}_\s \right)
\nonumber \\
&&+\ g_{12} \hat{\psi}^\dagger_1 \hat{\psi}^\dagger_2
\hat{\psi}_2 \hat{\psi}_1
- \alpha \left(\hat{\psi}^\dagger_1 \hat{\psi}^\dagger_1
\hat{\psi}_2 + {\rm h.c.} \right) \Bigg]
\label{hamiltonian}
\end{eqnarray}
where $\hat{\psi}^\dagger_\s({\bf x}),\hat{\psi}_\s({\bf x})$ are
bosonic field operators for atoms ($\s=1$) and molecules ($\s=2$),
$\hat{h}_{\s} = -(\hbar^2/2m_\s) \nabla^2 - \mu_\s$ are the
corresponding single particle Hamiltonians (focusing for
concreteness on the case of a homogeneous trap) with effective
chemical potentials $\mu_{1} = \mu$ and $\mu_2 = 2\mu - \nu$.
Chemical potential $\mu$ tunes the average {\em
total} number of atoms (whether free or bound into molecules) to
$N$, and detuning $\nu$ is related to the energy of a molecule at rest,
that can be experimentally controlled with a magnetic field. In
the dilute gas limit $g_1, g_2, g_{12}$ are proportional to the
2-body s-wave atom-atom, atom-molecule and molecule-molecule
scattering lengths, respectively, and $\alpha$ characterizes
\emph{coherent} atom-molecule interconversion rate,
encoding that molecules are composed of two atoms\cite{foot1}.

The mean-field phase diagram as a function of $\mu_{1,2}$ and
$\beta = 1/k_B T$ can be worked out by
minimizing the imaginary-time ($\tau$)
coherent-state action $S = \int_0^{\beta\hbar} d\tau \int d^d{\bf x}
\sum_{\s=1}^2 \left[\psi^*_\s\hbar \partial_\tau \psi_\s +
H_\mu(\psi^*_\s,\psi_\s) \right]$. Simple analysis leads to three
thermodynamically distinct phases (Fig.\ \ref{phasediagramA}): (i)
``normal'' (N): $\Psi_{10} \equiv \langle \hat \psi_1 \rangle =
0$, $\Psi_{20} \equiv \langle \hat \psi_2 \rangle = 0$, (ii)
``molecular superfluid'' (MSF): $\Psi_{10} = 0, \Psi_{20} \neq 0$,
(iii) ``atomic superfluid'' (ASF): $\Psi_{10} \neq 0, \Psi_{20}
\neq 0$. Condensed atoms cause $\alpha$ to act as an effective
field on the molecular order parameter $\Psi_{20}$, so an
equilibrium phase in which atoms are condensed, but molecules are
not, is forbidden \cite{comment1}.

We now examine in more detail these phases and corresponding phase
transitions.  Phase N is stable for $\mu_{1,2} < 0$, with $\mu$
determined by the total atom constraint $n = n_1 + 2 n_2$, which
in the non-interacting limit, appropriate to a dilute weakly
interacting gas, is given by:
\begin{equation}
n = \frac{1}{\Lambda_T^d} \left[f_{d/2}\left(e^{\beta\mu} \right)
+ 2^{(d+2)/2} f_{d/2}\left(e^{\beta(2\mu - \nu)} \right) \right],
\label{n_conserve}
\end{equation}
where $\Lambda_T = h/\sqrt{2\pi m_1 k_BT}$ is the thermal de
Broglie wavelength and $f_\alpha(z) = \sum_{n=1}^\infty
z^n/n^\alpha$ ($|z| < 1$) is the extended zeta-function.

The N-ASF transition line $T_{c1}(n,\nu)$ occurs at $\mu = 0$ for
$\nu > 0$, while the N-MSF line $T_{c2}(n,\nu)$ occurs at $\mu -
2\nu = 0$ for $\nu < 0$ (see Fig.\ \ref{phasediagramB}).  Using
the appropriate asymptotics of $f_\alpha(z)$, one obtains from
(\ref{n_conserve}):
\begin{equation}
T_{c\sigma}(n,\nu) =
\begin{cases}
T_{c0} \left[1 + a_\sigma \left(\frac{|\nu|}{k_B T_{c0}}
\right)^{\frac{d-2}{2}} \right],
& |\nu| \ll k_B T_{c0} \\
T_{c\sigma}^\infty = b_\sigma c^{2/d} T_{c0},
& |\nu| \gg k_B T_{c0},
\end{cases}
\label{Tc}
\end{equation}
with $c = 1 + 2^{(d+2)/2}$, $a_1 = 2^{(d+4)/2}
|\Gamma\left(\frac{2-d}{2}\right)|/d c\zeta(d/2)$, $a_2 = 2^{-d}
a_1$, $b_1 = 1$, $b_2 = 2^{-(d+2)/d}$  and $T_{c0} = (h^2/2\pi m_1
k_B) [n/c\zeta(d/2)]^{2/d}$ the transition temperature at the
tricritical point $\nu = 0$.

In the neighborhood of $T_{c1}$ the ``massive'' molecular field
$\hat{\psi}_2$ decouples at low energies (can be safely integrated out
of the partition function, leading to an effective quartic
coupling $g_1 \rightarrow \bar g_1 \equiv g_1 -
2\alpha^2/|\mu_2|$), and the N-ASF transition is identical to that
of a single-component system, continuous so long as $\bar g_1 >
0$. At $T=0$, the N-ASF transition takes place at vanishing atom
density (the N phase is simply a vacuum of atoms), and although
nontrivial, is exactly soluble \cite{uzunov}, corresponding to a
build-up of atomic superfluid (with condensate density $n_{10} =
|\Psi_{10}|^2 \sim |\mu|^{2\beta}$, with mean-field result $\beta
= 1/2$ for $d > 2$, and $\beta = d/4$ for $d < 2$) as the trap is
loaded. At $T \neq 0$ the N-ASF transition lies in the usual
$d$-dimensional XY-universality class \cite{foot2}.

Similarly, in the neighborhood of $T_{c2}$, $\hat \psi_1$
decouples and the resulting N-MSF transitions are in the same
universality classes discussed above.  The full phase boundary is
illustrated in Fig.\ref{phasediagramB}. In 3d it exhibits a
square-root singularity at the tricritical point and for $|\nu|
\to \infty$ asymptotes to the single-component BEC temperatures
$T_{c\sigma}^\infty$.

To study the MSF phase, we separate $\hat \psi_\sigma =
\Psi_{\sigma 0} + \hat{\phi}_\sigma$ into classical condensate
fields $\Psi_{\sigma 0}$ (with $\Psi_{10} = 0$ inside MSF) and
fluctuations about it. Within MSF it is sufficient to
expand $\hat H_\mu$ to second order in fluctuations $\hat
\phi_\sigma$, which leads to $\hat H_\mu = E^{(0)}(\Psi_{20}) +
\hat H^{(2)}$ with:
\begin{eqnarray}
E^{(0)} &=& \int d^d{\bf x}
\left(\Psi_{20}^* \hat h_2 \Psi_{20}
+ \frac{g_2}{2} |\Psi_{20}|^4 \right)
\\
{\hat H}^{(2)} &=& \int d^d{\bf x} \left[\sum_{\sigma = 1}^2
\hat \phi_\sigma^\dagger \tilde h_\sigma  \hat \phi_\sigma
+ \frac{1}{2} \left(\lambda_\sigma \hat \phi_\sigma^\dagger
\hat \phi_\sigma^\dagger + {\rm h.c} \right) \right],
\nonumber
\label{Htr}
\end{eqnarray}
in which $\tilde h_\sigma = \hat h_\sigma + r_\sigma$, 
$r_1 = g_{12} |\Psi_{20}|^2$, $r_2 = 2g_2 |\Psi_{20}|^2$, $\lambda_1 = -2\alpha
\Psi_{20}$, and $\lambda_2 = g_2 \Psi_{20}^2$.  The linear term in
$\hat \phi_2, \hat \phi_2^\dagger$ vanishes automatically by the
self-consistent choice of $\Psi_{20}$ as the true minimum of the
free energy. To lowest order this gives
\begin{equation}
n_{20} \equiv |\Psi_{20}|^2 = (2\mu - \nu)/g_2,
\label{n20}
\end{equation}
which coincides with the minimum of $E^{(0)}$ and allows us to
eliminate $\mu$ in favor of $\nu$ and $n_{20}$.

For a homogeneous system, $\hat{H}^{(2)}$ may be diagonalized by
Fourier transformation $\hat \phi_\sigma = V^{-1/2} \sum_{\kv}
e^{i\kv \cdot \xv} \hat a_{\sigma \kv}$, followed by independent
Bogoliubov transformations on atoms and molecules to new boson
operators $\gamma_{\s,\kv}, \gamma^\dagger_{\s,\kv}$:
\begin{eqnarray}
\hat{\gamma}_{\s,\kv} &=& u_{\s,\kv} \hat a_{\s,\kv} +
v_{\s,\kv} \hat a^\dagger_{\s,-\kv}
\nonumber \\
|u_{\s,\kv}|^2 &=& 1 + |v_{\s,\kv}|^2 = \frac{1}{2}
\left(\frac{\tilde \eps_{\s,\kv}}{E_{\s,\kv}} + 1 \right)
\nonumber \\
\hat{H}^{(2)} &=& \sum_{\sigma,\kv} E_{\s,\kv}
\left(\hat \gamma^\dagger_{\s,\kv} \hat \gamma_{\s,\kv}
- |v_{\s,\kv}|^2 \right),
\end{eqnarray}
in which $\tilde \eps_{\s,\kv} = \eps_{\s,\kv} - \mu_\s + r_\s$,
$\eps_{\s,\kv} = \hbar^2k^2/2m_\sigma$ and $E_{\s,\kv} =
\sqrt{\tilde \eps_{\s,\kv}^2 - |\lambda_\sigma|^2}$, and
$v_{\s,\kv},u_{\s,\kv}^*$ have the same phase as $\lambda_\sigma$.
Using (\ref{n20}) one obtains
\begin{eqnarray}
E_{1,\kv} &=& \sqrt{\left[\eps_{1,\kv} - \nu/2
- (g_2/2 - g_{12}) n_{20} \right]^2 - 4 \alpha^2 n_{20}}\;,
\nonumber \\
E_{2,\kv} &=& \sqrt{\eps_{2,\kv}^2 + 2g_2 n_{20}\, \eps_{2,\kv}}\;.
\end{eqnarray}
The MSF ground state is defined as vacuum of atomic and molecular
Bogoliubov quasi-particles $\gamma_{\s,\kv}|{\rm MSF}\rangle = 0$.
It is easy to show that it is given by:
\begin{equation}
|{\rm MSF}\rangle=e^{\Psi_{20} \hat a_{2,0}^\dagger}
\prod_{\s,\kv} e^{-\chi_{\s,\kv} \hat a_{\s,\kv}^\dagger
\hat a_{\s,\kv}^\dagger}|0\rangle,
\end{equation}
which is a coherent state in the $\kv = 0$ molecular mode and
shows BCS-like pairing correlations between time-reversed ($\kv,
-\kv$) atomic and molecular single particle states.  The amplitude
$\chi_{\s,\kv} =  v_{\s,\kv}/u_{\s,\kv}$ is the Fourier transform
of the wavefunction of $\kv \neq 0$ atom ($\sigma=1$) and molecule
($\sigma=2$) pairs.

As in a single-component SF the relation (\ref{n20}) (and more
fundamentally, the Goldstone theorem \cite{ZinnJustin}) ensures
that MSF exhibits a gapless sound mode $E_{2,\kv} \approx \hbar
v_0 k$, for small $k$, with $v_0 = \sqrt{g_2 n_{20}/2m_1}$,
corresponding to collective, long wavelength oscillations of the
molecular condensate. The resulting $|\kv|$ singularity in
$\chi_{2,\kv}$ induces a long ranged power law tail in the Fourier
transform $\chi_2(r) \sim 1/r^{d+1}$.  In addition, we find a
\emph{gapped} excitation branch $E_{1,\kv}$ describing atomic-like
excitations, whose spectrum and eigenmodes ($\hat{\gamma}_{1,\kv},
\hat{\gamma}_{1,\kv}^\dagger$ do not even carry a definite atom
number) are qualitatively distinct from unpaired atomic
excitations, $\hat{a}_{1,\kv}, \hat{a}_{1,\kv}^\dagger$ of the N
phase. The condition of being in the MSF is that the atomic gap
$E^{(1)}_{\rm gap} = \sqrt{\eps_+ \eps_-}$, where
\begin{equation}
\eps_{\pm} \equiv -\nu/2-(g_2/2 - g_{12})n_{20}
\pm 2\alpha \sqrt{n_{20}},
\label{gap}
\end{equation}
remains positive.  Correspondingly, the atomic pair wavefunction
$\chi_1(r) \sim e^{-r/\xi_1}$ decays exponentially at large
distance.  The correlation length $\xi_1 = \hbar/\sqrt{2m \eps_-}$
characterizes atom-pair size, and diverges as
$E^{(1)}_{\rm gap}$ vanishes.  The MSF-ASF transition takes place when
the size of atom-pairs becomes comparable to the intermolecular
separation, and $\xi_1$ is the diverging size of the coherent
exchange loops of atoms between overlapping pairs.

In the present quadratic approximation, the free energy is given
by
\begin{eqnarray}
f &=& -\mu_2|\Psi_{20}|^2 + \frac{g_2}{2}|\Psi_{20}|^4
- \frac{1}{2V} \sum_{\s,\kv}(\tilde{\eps}_{\s,\kv} - E_{\s,\kv})
\nonumber\\
&&+\ \frac{k_B T}{V}\sum_{\s,\kv}\ln(1-e^{-E_{\s,\kv}/k_BT}).
\end{eqnarray}
The condition of fixed density $n = -(\partial f/\partial
\mu)_{n_{20}}$,
\begin{equation}
n = 2n_{20} + \frac{1}{V} \sum_{\s,\kv} \s
\left(v_{\s,\kv}^2 + \frac{u_{\s,\kv}^2 + v_{\s,\kv}^2}
{e^{E_{\s,\kv}/{k_BT}}-1} \right),
\end{equation}
where the first term under the summation represents the $T=0$
interaction-induced condensate depletion, may be used to determine
the condensate density $n_{20}(n,\nu,T)$.  At $T=0$, in the
weakly-interacting limit ($a^3 n\ll 1$), in 3d we find:
\begin{eqnarray}
n_{20} &=& \frac{1}{2} n - \frac{2^{(d-2)/2}
\Gamma\left(\frac{d-1}{2}\right)
\Gamma\left(\frac{4-d}{2}\right)}{d \sqrt{\pi}
\Gamma\left(\frac{d}{2}\right)} (n a_2)^{d/2}
\nonumber \\
&&-\ \frac{[(n_\nu - n) \bar a_{12}]^{d/2}}
{2^{d+1} \Gamma\left(\frac{d}{2}\right)}
{\cal I}_d\left(\frac{\bar \alpha \sqrt{n}}{n_\nu - n} \right)
\label{n20n}
\end{eqnarray}
where $\bar a_{12} \equiv a_{2} - 2 a_{12}$,
$a_{2}, a_{12}$ are molecule-molecule and atom-molecule
scattering lengths, respectively, 
defined by $g_{12,2} = 4\pi\hbar^2 a_{12,2}/m_2$,
the ``detuning density'' is $n_\nu = m_1 |\nu|/(\pi \hbar^2 \bar
a_{12})$, $\bar \alpha = 2^{3/2} m_1 |\alpha|/(\pi\hbar^2 \bar
a_{12})$, and ${\cal I}(y) = y^2 \int_0^\infty dx
x^{d/2}[(1+x)^2 - y^2]^{-3/2}$ is a scaling function describing
additional molecular-condensate depletion due to atom pairs.  Equation
(\ref{n20n}), together with the vanishing of $\eps_-$, determines the
MSF-ASF phase boundary. At $T=0$, approximating $n_{20} \approx n/2$,
one finds:
\begin{equation}
\nu_c(n) \approx -(g_2/2-g_{12})n - 2\alpha\sqrt{2n},
\label{nuc}
\end{equation}
which is illustrated for $g_2 > 2g_{12}$ in Fig.\ \ref{phasediagramA}.
At finite $T$, in the noninteracting limit, $v_{\s,\kv} = 0$ and
$u_{\s,\kv} = 1$, molecular condensate density reduces to
\begin{equation}
n_{20}(\nu,T) = \frac{n}{2} \left[1 -
\left(\frac{T}{T_{c2}^\infty} \right)^{3/2} \right]
- \frac{1}{2 \Lambda_T^3} f_{3/2}\left(e^{-|\nu|/2k_B T} \right),
\end{equation}
with the standard BEC result (first term) corrected by depletion
due to thermal atomic excitations (last term), that is exponentially
small at large $|\nu|$ and low $T$.

From the structure of the interconversion ($\alpha$) term in $\hat
H_\mu$ it is clear that it is a $Z_2$ (Ising) symmetry that is
broken at the MSF$\rightarrow$ASF transition \cite{comment2}. We
now show that for a homogeneous trap, both the $T=0$ and finite
$T$ MSF-ASF transitions lie in the Ising universality class. This
can be most easily seen from the coherent state action, which,
when expressed in terms of real and imaginary parts of the atomic
field $\hat \psi_1 = \hat \psi_R + i \hat \psi_I$, in MSF phase
reduces to:
\begin{eqnarray}
S &=& \int_0^{\beta\hbar} d\tau \int d^d{\bf x}
\bigg[-2i \psi_I\hbar \partial_\tau\psi_R
+ \psi_R(\hat{h}_1 - 2\alpha\Psi_{20})\psi_R
\nonumber \\
&&\ \ \ \ \ \ \ \ \ \
+\ \psi_I(\hat{h}_1 + 2\alpha\Psi_{20})\psi_I\bigg]
+ S_{\rm int}
\end{eqnarray}
where $S_{\rm int}$ are terms not essential for our argument. In
this form it is clear that in the presence of the molecular
condensate, $|\Psi_{20}| > 0$, positive $\alpha$ reduces the
$O(2)$ symmetry to $Z_2$, with $\psi_R$ reaching criticality {\em
before} $\psi_I$.  Because canonically conjugate field $\psi_I$
remains ``massive'' at the critical point [defined by where the
coefficient of $\psi_R^2$ vanishes, consistent with (\ref{nuc})],
it can be safely integrated out and leads to a $d+1$-dimensional
(Lorentz-invariant) action even in the scalar order parameter
$\psi_R$. Therefore, as asserted above, the $T=0$ MSF-ASF
transition is in the $(d+1)$-dimensional Ising universality class.
The Ising transition is well studied, and leads to the following
predictions \cite{ZinnJustin}. For $d=3$, up to logarithmic
corrections, the mean-field theory derived above will be accurate.
On the other hand in 2d, MSF-ASF exponents are nontrivial but are
well-known. For example, standard scaling arguments predict:
\begin{eqnarray}
n_{10} \sim |\nu - \nu_c|^{2\beta_I},\ \
E_{gap}^{(1)} \sim |\nu - \nu_c|^{z_I \nu_I},
\end{eqnarray}
where $\beta_I \approx 0.31$, $z_I = 1$, and $\nu_I \approx 0.63$
are 3d Ising exponents. These, together with the
relevance of $T$ at this quantum critical point also imply a {\em
universal} shape of the MSF-ASF phase boundary $\nu_c(n,T) \sim
\nu_c(n,0) + a\, T^{1/\nu_I}$ in Fig.\ \ref{phasediagramB}. One
may hope that when long-lived molecular condensates are produced,
nontrivial behavior of $E_{gap}^{(1)}(\nu)$ maybe observed in
Ramsey fringes experiments \cite{donley}.

Although from symmetry point of view ASF state is quite
conventional, it also exhibits a set of interesting features not
found in a single-component SF.  For example, because it is a
discrete (Ising) symmetry that distinguishes ASF from MSF, it will
exhibit a gapped (Ising) mode in addition to the gapless
second-sound mode. As discussed above \cite{comment1} for $\alpha
\neq 0$, finite $\Psi_{20}$ is always induced within the ASF.
However, in the limit of small $\alpha$, we predict a sharp
crossover (indicated by a dashed $\mu_2^{\rm eff} \equiv \mu_2 -
\mu_1 g_{12}/\bar g_1 = 0$ curve in Figs.\ \ref{phasediagramA},
\ref{phasediagramB}, where the $\psi_2$ field would order for
$\alpha = 0$\ \cite{comment1}) 
between small molecular condensate $n_{20} \approx
(\alpha \mu_1/g_1 \mu_2^{\rm eff})^2$ for $\mu_2^{\rm eff} < 0$
and large $n_{20} \approx \mu_2^{\rm eff}/g_2$ for $\mu_2^{\rm
eff} > 0$, and  vanishes as $n_{20} \sim (\alpha
n_{10})^{2/\delta}$ for $\mu_2^{\rm eff} = 0$ and weak $\alpha$,
with $\delta$ a universal XY-model magnetic field exponent
($\delta_{d=3} \simeq 5$) \cite{ZinnJustin}.

\begin{figure}[bth]
\centering
\setlength{\unitlength}{1mm}
\begin{picture}(70,22)(0,0)
\put(-35,-92){\begin{picture}(50,20)(0,0)
\includegraphics{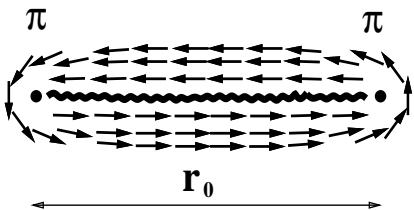}
\end{picture}}
\end{picture}
\caption{$2\pi$ atomic condensate vortex in ASF splits into
$\pi+\pi$ vortex pair connected by a ``normal'' domain wall.}
\label{vortex_pi}
\end{figure}

Another experimentally interesting feature of ASF is that a
seemingly standard $2\pi$ vortex in atomic condensate $\Psi_{10}$
will generically split into two $\pi$ vortices (see Fig.\
\ref{vortex_pi}) confined by a domain wall of length
\begin{equation}
r_0 = \sqrt{\hbar^2 n_{20}^{3/2}/2m\alpha n_{10}^2},
\end{equation}
that diverges as ASF$\rightarrow$MSF phase boundary is approached
\cite{LRunpublished}.
This arises because in large $\alpha$ limit a $2\pi$ vortex in atomic
condensate induces a $4\pi$ vortex in a molecular condensate. Such
double molecular vortex is unstable to two fundamental $2\pi$
molecular vortices, that, in 2d repel logarithmically, but are confined 
linearly inside the ASF phase.

Finally, standard renormalization-group analysis shows that
upper-critical dimensions for the tricritical point (dominated by
scaling of $\alpha$) are $d_{uc}^{T > 0}=6$ and $d_{uc}^{T=0} =
4$, and therefore should display nontrivial critical properties
\cite{LRunpublished}.

We plan to present analysis of these problems as well as
generalization to experimentally relevant harmonic trap in a future
publication.

To conclude, we have studied a mixture of bosonic atoms and their
diatomic molecules and predicted an existence of a molecular
superfluid phase, qualitatively distinct from the normal and atomic
superfluid states. Ising transition MSF-ASF should be observable in
such systems as a function of detuning $\nu$, temperature and/or
density.

We acknowledge support by NSF DMR-0321848 and Packard Foundation
(LR,JP), and thank Chris Greene for discussions and Steve Girvin for 
comments on the manuscript.


\vspace{-0.5cm}
\begin{thebibliography}{99}
\vspace{-0.5cm}

\bibitem{JILAbec} M. H. Anderson, {\em et al.}, {\em Science} {\bf
269}, 198 (1995).

\bibitem{MITbec} K. B. Davis {\em et al.}, \prl {\bf 75}, 3969
(1995).

\bibitem{donley} E. A. Donley, {\em et al.}, {\em Nature}
(London) {\bf 417}, 529 (2002).

\bibitem{jinBCS} C. A. Regal, {\em et al.}, {\em Nature} {\bf 424}, 47 (2003).

\bibitem{foot1} The effective Hamiltonian (\ref{hamiltonian}) may
be derived in various limits from an underlying atomic model, in
which molecules are explicitly identified as bound or quasibound
states of atom pairs (including atomic spin dynamics, if
necessary); J. I. Park, L. Radzihovsky and P. B. Weichman, manuscript 
in preparation.

\bibitem{comment1} This contrasts with a mixture of two {\em
independent} species of bosons with $\alpha=0$, that in addition
admits a $\Psi_{10} \neq 0$, $\Psi_{20} = 0$ distinct phase.

\bibitem{foot2} In the dilute, weakly interacting BEC limit the
critical behavior will be that of the ideal gas (i.e., Gaussian),
crossing over to XY behavior only very close to $T_{c1}$:  P. B.
Weichman, M. Rasolt, M. J. Stephen and M. E. Fisher, Phys.\ Rev.\
B {\bf 33}, 4632 (1986);  P. B. Weichman, Phys.\ Rev.\ B {\bf 38},
8739 (1988).

\bibitem{GradshteynRizhik} I. S. Gradshteyn and I. M. Rizhik, {\em
Table of Integrals, Series and Products} (Academic Press, New
York, 1980).

\bibitem{uzunov} D. I. Uzunov, Phys.\ Lett.\ {\bf 87A}, 11 (1981).


\bibitem{ZinnJustin} J. Zinn-Justin, {\em Quantum Field Theory and
Critical Phenomena}, (Oxford Science Puplications, Oxford, 1989).

\bibitem{comment2} In the ``hard-spin'' $\psi_\s=e^{i\theta_\s}$
description, interconversion term reduces to $\alpha
\cos(2\theta_1-\theta_2)$, locking atomic phase to the phase of
the molecular condensate up to $\pi$, corresponding to the
unbroken $Z_2$ symmetry of the MSF. In terms of a vector order
parameter, MSF is characterized by a finite quadrapole moment
($l=2$ angular harmonic, akin to nematic order in liquid
crystals), and ASF by a vector order parameter ($l=1$ harmonic).

\bibitem{LRunpublished} L. Radzihovsky, unpublished.

\end{thebibliography}
\end{document}